\documentclass[runningheads]{llncs}
\usepackage{graphicx}
\usepackage{hyperref}
\usepackage{amsmath}
\usepackage{multirow}
\usepackage{color,soul}
\usepackage{bm}

\begin{document}
\title{Adversarial optimization for joint registration and segmentation in prostate CT radiotherapy}
\titlerunning{Adversarial optimization for joint registration and segmentation} 

\author{Mohamed S. Elmahdy\inst{1}, Jelmer M. Wolterink\inst{2}, Hessam Sokooti \inst{1}, Ivana I\v{s}gum\inst{2} \and Marius Staring\inst{1,3}}

\authorrunning{M. Elmahdy et al.}

\institute{Division of Image Processing, Department of Radiology, Leiden University Medical Center, 2300 RC  Leiden, The Netherlands \\ \email{m.s.e.elmahdy@lumc.nl} \and
Image Sciences Institute, University Medical Center Utrecht, Utrecht, The Netherlands \and
Department of Radiation Oncology, Leiden University Medical Center, 2300 RC  Leiden, The Netherlands}
\maketitle              
\begin{abstract}
Joint image registration and segmentation has long been an active area of research in medical imaging. Here, we reformulate this problem in a deep learning setting using adversarial learning. We consider the case in which fixed and moving images as well as their segmentations are available for training, while segmentations are not available during testing; a common scenario in radiotherapy. The proposed framework consists of a 3D end-to-end generator network that estimates the deformation vector field (DVF) between fixed and moving images in an unsupervised fashion and applies this DVF to the moving image and its segmentation. A discriminator network is trained to evaluate how well the moving image and segmentation align with the fixed image and segmentation. The proposed network was trained and evaluated on follow-up prostate CT scans for image-guided radiotherapy, where the planning CT contours are propagated to the daily CT images using the estimated DVF. A quantitative comparison with conventional registration using \texttt{elastix} showed that the proposed method improved performance and substantially reduced computation time, thus enabling real-time contour propagation necessary for online-adaptive radiotherapy.

\keywords{Deformable image registration, adversarial training, image segmentation, contour propagation, radiotherapy}
\end{abstract}

\section{Introduction}
Joint image registration and segmentation (JRS) has long been an active area of research in medical imaging. Image registration and segmentation are closely related and complimentary in applications such as contour propagation, disease monitoring, and data fusion from different modalities. Image registration could be enhanced and improved using an accurate segmentation, and vice versa registration algorithms could be used to improve image segmentation.

An important application in which coupling of image registration and segmentation is crucial, is online adaptive image-guided radiotherapy. In this application, clinically approved contours are propagated from an initial \textit{planning} CT scan to \textit{daily} inter-fraction CT scans of the same patient. Image registration can be used to correct for anatomical variations in shape and position of the underlying organs, as well as to compensate for any misalignment in patient setup. Ideally, contours should be propagated quickly to allow immediate computation of a new dose distribution. With these propagated contours, margins can be smaller and treatment-related complications may be reduced. Thus, it is important that the daily contours are of high quality, are consistent with the planning contours, and are generated in near real-time.

In the last decade, researchers have been working on fusing image registration and segmentation. Lu \emph{et al.} \cite{Lu} proposed a Bayesian framework for modelling segmentation and registration such that these could alternatingly constrain each other. Yezzi \emph{et al.} \cite{Yezzi} proposed using active contours to register and segment images. Unal \emph{et al.} \cite{Unal}, generalizing on \cite{Yezzi}, proposed to use partial differential equations without any shape prior. Most of these methods require long computation times and complex parameter tuning. Recently, the widespread adoption of deep learning techniques has led to remarkable achievements in the field of medical imaging \cite{Litjens}. Among these techniques are generative adversarial networks (GANs), which are defined by joint optimization of a generator and discriminator network \cite{Goodfellow}. GANs have boosted the performance of traditional networks for image segmentation \cite{Kazeminia} as well as registration \cite{Haskins}. Recently, Mahapatra \emph{et al.} \cite{Mahapatra} proposed a GAN for joint registration and segmentation of 2D chest X-ray images. However, this method requires reference deformation vector fields (DVFs) for training. In practice, these are often unavailable and it may be more practical to perform unsupervised registration \cite{Vos}, i.e. training without reference DVFs.

In this paper, we introduce a fast unsupervised 3D GAN to jointly perform deformable image registration and segmentation. A generator network estimates the DVF between two images, while a discriminator network is trained simultaneously to evaluate the quality of the registration and the segmentation and propagate the feedback to the generator network. We consider the use-case in which fixed and moving images as well as their segmentations are available for training, which is a common scenario in radiation therapy. However, no segmentations are required for DVF estimation during testing. This paper has the following contributions. First, we propose an end-to-end 3D network architecture, which is trained in an adversarial manner for joint image registration and segmentation. Second, we propose a strategy to generate well-aligned pairs to train the discriminator network with. Third, we leverage PatchGAN as a local quality measure of image alignment. Fourth, the proposed network is much faster and more accurate than conventional registration methods. 

We quantitatively evaluate the proposed method on a prostate CT database, which shows that the method compares favorably to \texttt{elastix} software \cite{Klein}.

\section{Methods}
Image registration is the transformation of a moving image $I_m$ to the coordinate system of a fixed image $I_f$. In this paper, we assume that all image pairs are affinely registered beforehand, and we focus on local non-linear deformations. 

In conventional contour propagation algorithms, registration and segmentation are disjoint. First, the DVF $\Phi$ is estimated using image registration, and then $\Phi$ is used to warp the contours $S_m$ to the fixed coordinate space. Afterwards, during system evaluation, a similarity measure such as the Dice similarity coefficient (DSC) can be used to measure the quality of the propagated contours w.r.t. ground truth contours, but this information is not fed back to the registration algorithm. We call this an \textit{open loop} system. In contrast, this paper proposes an end-to-end \textit{closed loop} system to improve image registration based on feedback on the registration as well as the segmentation quality.

\subsection{Adversarial Training}
We propose to train a GAN containing two CNNs: a generator network that predicts the DVF $\Phi$ given $I_f$ and $I_m$, and a discriminator network that assesses the alignment of $I_f(\bm{x})$ and $I_m(\Phi(\bm{x}))$ as well as the overlap between $S_f(\bm{x})$ and $S_m(\Phi(\bm{x}))$. Hence, we assume that $S_f$ and $S_m$ are both available, but during training only. The GAN is trained using a Wasserstein objective \cite{Arjovsky}, which has empirically been shown to improve training stability and convergence compared to the GAN objective in \cite{Goodfellow}. Equations (\ref{eq:2}) and (\ref{eq:3}) list the generator loss $L_G^{GAN}$ and the discriminator loss $L_D^{GAN}$ of WGAN:
\begin{align}
L_G^{GAN} &=  E\left[ D(I_f(\bm{x}), I_m(\Phi(\bm{x})), S_m(\Phi(\bm{x})))\right],
\label{eq:2}\\
L_D^{GAN} &=  E\left[ D(I_f(\bm{x}), I_m(\Phi(\bm{x})), S_m(\Phi(\bm{x})))\right] - \left[ D(I_f, \Theta(I_f), S_f) \right],
\label{eq:3}
\end{align}
where $G$ and $D$ denote the generator and discriminator networks with trainable parameters and $\Phi$ is the DVF provided by $G$. In a GAN, the discriminator is trained to distinguish between \textit{real} and \textit{fake} samples. In this case, fake samples are the triple $(I_f, I_m(\Phi), S_m(\Phi))$, while real samples should be well-aligned images. As we perform unsupervised registration, and assume no knowledge about the ideal alignment of two images, we synthesize such image based on the fixed image and its segmentation alone: $(I_f, \Theta(I_f), S_f)$. Hence, $\Theta$ in Equation (\ref{eq:3}) is a random combination of disturbance functions, as follows. First, to mimic imaging noise, Gaussian noise and Gaussian smoothing are added with zero mean and a standard deviation of 0.04. Second, to mimic contrast variations, we apply gamma correction with a random gamma factor in the range $[-0.4, 0.4]$. Third, we mimic interpolation errors by applying a random deformation of less than 0.5 mm and resample the images using that deformation using linear interpolation.

In addition to these image-based quality measures, we include the segmentation of the deformed moving image as input to the discriminator in order to enforce DVFs that are consistent with the moving segmentation. We test two designs. The first design concatenates the segmentation as a third input channel in the discriminator, next to the fixed and moving image channels. The second design multiplies the fixed and moving image channel with the corresponding segmentation, so that the network learns to focus on the target structures and organs-at-risk instead of on the bowels and other less relevant soft tissue. These designs are named JRS-GAN$^a$ and JRS-GAN$^b$, respectively. 

We found that training the network using WGAN loss only, resulted in slow convergence and suboptimal registrations. Thus, a similarity loss $L_{sim}$, based on image similarity and segmentation overlap, was added to the generator:
 
\begin{gather}
L_{sim} =  (1 - \mathrm{DSC}(S_m(\Phi(\bm{x})), S_f(\bm{x}))) + (1 - \mathrm{NCC}( I_m(\Phi(\bm{x})), I_f(\bm{x}))),
\label{eq:4}
\end{gather}
where DSC is the Dice similarity coefficient and NCC is normalized cross-correlation. Adding the DSC to $L_{sim}$ ensures that the registration improves the segmentation and vice versa. Furthermore, to ensure smooth and continuous DVFs, the bending energy penalty of the DVF, $L_{smooth}$, was added as a regularization term to the overall generator loss, which was defined as:
\begin{gather}
L_G = L_{sim} +\lambda_1 L_{smooth} +  \lambda_2 L_G^{GAN},
\label{eq:6}
\end{gather}
where $\lambda_1$ and $\lambda_2$ are weights for the DVF smoothness and the generator loss. 

During training of the network, for every iteration of the generator we used 100 iterations of the discriminator, for the first 25 iterations. After that we used the ratio 1:5. In each iteration, weights of the discriminator were clipped to the range $[-0.01, 0.01]$ \cite{Arjovsky}.

\subsection{Network Architectures}
\label{sec:architectures}
\begin{figure*}[t!]
	\centering
	\includegraphics[width=\linewidth]{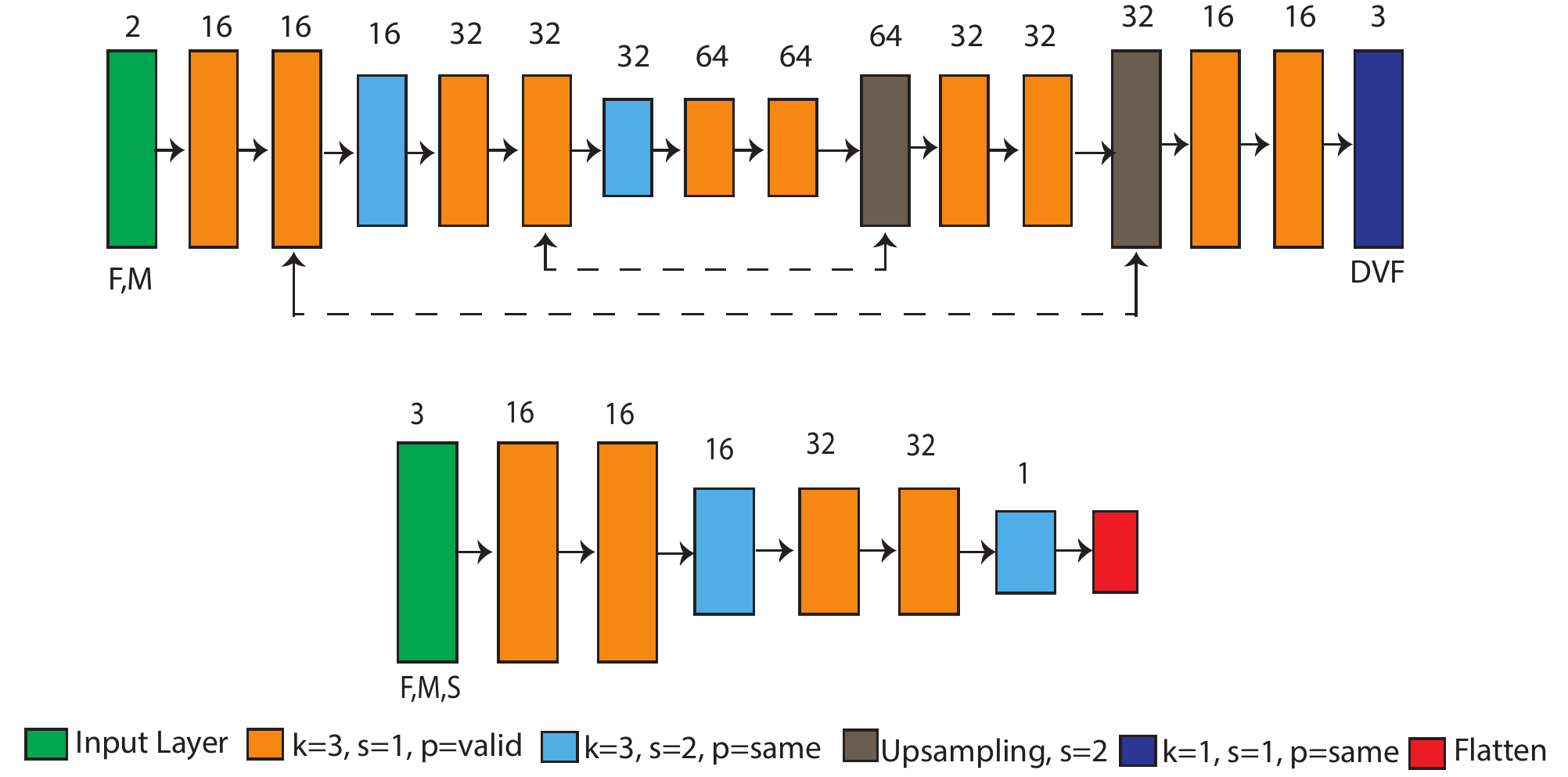} 
	\caption{The proposed generator (top) and discriminator (bottom) networks, where k, s, and p represent the kernel size, stride size, and padding option, respectively. The numbers above the different layers represent the feature maps.}
	\label{fig:cnn_network}
\end{figure*}

\noindent\textbf{Generator Network} To estimate the parametric mapping function $\Phi$ between the fixed and moving images we use a 3D network similar to the U-net \cite{Ronneberger}. Figure~\ref{fig:cnn_network} shows the network design in more detail. The input to the network is the concatenation of $I_f$ and $I_m$. The network encodes the image pairs through a set of 3$\times$3$\times$3 convolution layers followed by LeakyReLU and batch normalization layers. Strided convolutions are used in the contractive path and upsampling layers are used in the expanding path. The output size of the network is smaller than the input size in order to consider a larger field of view. A resampling network adopted from NiftyNet \cite{Gibson} is used to warp the images using the estimated DVF during training time so that the network can be trained end-to-end. 

\noindent\textbf{Discriminator Network} 
The discriminator is responsible for assessing whether the image pairs are well-aligned or not, as well as assessing whether the segmentations overlap. Figure~\ref{fig:cnn_network} shows the network design, which is similar to the contracting path of the generator. The discriminator network was trained using PatchGAN \cite{Isola}. Hence, instead of representing the quality of the whole patch with a single number, the network could quantify the sub-patch quality locally.

. 

\section{Experiments and Results}

\subsection{Dataset, evaluation criteria and implementation details}
This study includes eighteen patients who underwent intensity-modulated radiation therapy for prostate cancer in 2007 at Haukeland university hospital \cite{Muren}. Each patient had a planning CT as well as 7 to 10 inter-fraction repeat CT scans. The prostate, lymph nodes, seminal vesicles, as well as the rectum and bladder were annotated. Each scan has 90 to 180 slices with a slice thickness of around 2 to 3 mm. All the slices were of size 512 $\times$ 512 with an in-plane resolution of around 0.9 mm. All the volumes were affinely registered using \texttt{elastix}. The volumes were resampled to isotropic voxel size of 1$\times$1$\times$1 mm. All volumes intensities were scaled to [-1, 1]. We split the dataset into 111 image pairs (from 12 patients) for training and validation and 50 image pairs (6 patients) for testing. 

The quality of registration is quantified geometrically in 3D by comparing the manual delineations of the daily CT with the automatically propagated contours. We use the mean surface distance (MSD), and the 95\% Hausdorff distance (HD). A Wilcoxon signed rank test at $p = 0.05$ is used to compare results.

The networks were implemented using TensorFlow (version 1.13) \cite{Abadi} with the RMSProp optimizer using a learning rate of $10^{-5}$. The networks were trained and tested on an NVIDIA Tesla V100 GPU with 16 GB of memory. From each image pair, 1000 patches of size 96$\times$96$\times$96 voxels were sampled within the torso mask. To improve stability, the network was trained to warp the fixed patch to the moving patch and vice versa at the same training iteration. The magnitude of the three loss terms in Equation (\ref{eq:4}) was scaled by setting $\lambda_1 = 1$ and $\lambda_2 = 0.01$.

\subsection{Experiments and results}

Tables \ref{table:msd} and \ref{table:hd} provide quantitative results comparing the following methods. First, we include conventional iterative methods using \texttt{elastix} software \cite{Klein} with NCC (\texttt{elastix}-NCC) and MI (\texttt{elastix}-MI) similarity measures, using the settings from \cite{Yuchuan}. Second, we evaluate two unsupervised deep learning-based methods without adversarial feedback: One uses the generator trained with the NCC loss (Reg-CNN), similar to \cite{Vos}; the other uses the generator with both the NCC and DSC loss (JRS-CNN). Third, we evaluate several versions of our GAN-based approach. To study the effect of adversarial training without added segmentations, we perform an experiment named Reg-GAN. Finally, we evaluate the proposed JRS-GAN$^a$ and JRS-GAN$^b$ methods.

\begin{table}[t]
	\centering
	\setlength{\tabcolsep}{3pt}
	\caption[Table caption text]{MSD (mm) values for different experiments, where $\dagger$ and $\ddagger$ represent a significant difference compared to \texttt{elastix}-MI and Reg-CNN, respectively.}
	\resizebox{\textwidth}{!}{
	\begin{tabular}{llllll} 
		&Prostate&Seminal vesicles&Lymph nodes&Rectum& Bladder \\ \hline
		Evaluation & $\mu \pm \sigma$ & $\mu \pm \sigma$ & $\mu \pm \sigma$ & $\mu \pm \sigma$ & $\mu \pm \sigma$ \\ \hline
		\texttt{elastix}-NCC & $1.81 \pm 0.7$ & $2.80 \pm 1.6$ & $1.19 \pm 0.4$ & $3.79 \pm 1.2$ &  $5.31 \pm 2.6$  \\ 
		\texttt{elastix}-MI & $1.73 \pm 0.7$ & $2.70 \pm 1.6$ & $1.18 \pm 0.4$ & $3.68 \pm 1.2$ &  $5.26 \pm 2.6$  \\  \hline
		Reg-CNN& $1.44 \pm 0.5^\dagger$ & $2.09 \pm 1.7^\dagger$ & $1.22 \pm 0.3$ & $2.59 \pm 1.3^\dagger$ &  $4.18 \pm 2.6^\dagger$  \\ 
		JRS-CNN& $1.18 \pm 0.4^{\dagger \ddagger}$ & $1.91 \pm 1.6^{\dagger \ddagger}$ & $1.02 \pm 0.3^{\dagger \ddagger}$ & $2.32 \pm 1.3^{\dagger \ddagger}$ &  $2.37 \pm 2.0^{\dagger \ddagger}$  \\ 
		\hline
		Reg-GAN& $1.40 \pm 0.5^\dagger$ & $2.14 \pm 1.7^\dagger$ & $1.06 \pm 0.3^{\dagger \ddagger}$ & $2.72 \pm 1.3^\dagger$ &  $4.31 \pm 2.8^\dagger$  \\ 
		JRS-GAN$ ^a $& $\mathbf{1.13 \pm 0.4}^{\dagger \ddagger}$ & $\mathbf{1.81 \pm 1.6}^{\dagger \ddagger}$ & $\mathbf{1.00 \pm 0.3 }^{\dagger \ddagger}$ & $\mathbf{2.21 \pm 1.3}^{\dagger \ddagger}$ &  $\mathbf{2.29 \pm 2.0}^{\dagger \ddagger}$  \\ 
		JRS-GAN$ ^b $& $1.17 \pm 0.4^{\dagger \ddagger}$ & $1.90 \pm 1.5^{\dagger \ddagger}$ & $1.01 \pm 0.3^{\dagger \ddagger}$ & $2.34 \pm 1.3^{\dagger \ddagger}$ &  $2.41 \pm 2.1^{\dagger \ddagger}$  \\ \hline
	\end{tabular}
	}
	\label{table:msd} 
\end{table}

\begin{table}[t]
	\centering
	\setlength{\tabcolsep}{3pt}
	\caption[Table caption text]{\%95HD (mm) values for different experiments, where $\dagger$ and $\ddagger$ represent a significant difference compared to \texttt{elastix}-MI and Reg-CNN, respectively.}
	\resizebox{\textwidth}{!}{
	\begin{tabular}{llllll} 
		&Prostate&Seminal vesicles&Lymph nodes&Rectum& Bladder \\ \hline
		Evaluation & $\mu \pm \sigma$ & $\mu \pm \sigma$ & $\mu \pm \sigma$ & $\mu \pm \sigma$ & $\mu \pm \sigma$ \\ \hline
		\texttt{elastix}-NCC & $4.2 \pm 1.8$ & $6.1 \pm 3.3$ & $\mathbf{2.8 \pm 1.0}^\ddagger$ & $11.0 \pm 5.2$ &  $15.4 \pm 8.4^\ddagger$  \\ 
		\texttt{elastix}-MI & $4.0 \pm 1.7$ & $6.0 \pm 3.7$ & $2.8 \pm 1.0^\ddagger$ & $10.9 \pm 5.2$ &  $15.3 \pm 8.3^\ddagger$  \\  \hline
		Reg-CNN& $5.3 \pm 2.5$ & $6.2 \pm 3.5$ & $4.4 \pm 1.4$ & $11.0 \pm 6.5$ &  $16.6 \pm 9.3$  \\ 
		JRS-CNN& $3.6 \pm 1.5^{\dagger \ddagger}$ & $5.4 \pm 3.4^{\dagger \ddagger}$ & $3.1 \pm 0.9{\ddagger}$ & $10.3 \pm 6.7^{\dagger \ddagger}$ &  $11.6 \pm 10.5^{\dagger \ddagger}$  \\ 
		\hline 
		Reg-GAN& $4.3 \pm 2.1{\ddagger}$ & $6.0 \pm 3.6$ & $3.4 \pm 1.0{\ddagger}$ & $11.1 \pm 6.4$ &  $16.2 \pm 9.6{\ddagger}$  \\
		JRS-GAN$ ^a $& $\mathbf{3.4 \pm 1.4}^{\dagger \ddagger}$ & $\mathbf{5.3 \pm 3.3}^{\dagger \ddagger}$ & $3.1 \pm 0.9 {\ddagger}$ & $\mathbf{10.0 \pm 6.7}^{\dagger \ddagger}$ &  $\mathbf{11.0 \pm 10.3}^{\dagger \ddagger}$  \\ 
		JRS-GAN$ ^b $& $3.5 \pm 1.4^{\dagger \ddagger}$ & $5.6 \pm 3.7{\ddagger}$ & $3.0 \pm 1.0{\ddagger}$ & $10.5 \pm 6.8^{\dagger \ddagger}$ &  $11.4 \pm 10.6^{\dagger \ddagger}$  \\ \hline
	\end{tabular}
	}
	\label{table:hd}
\end{table}

\begin{figure}[t]
	\centering
	\includegraphics[width=\linewidth]{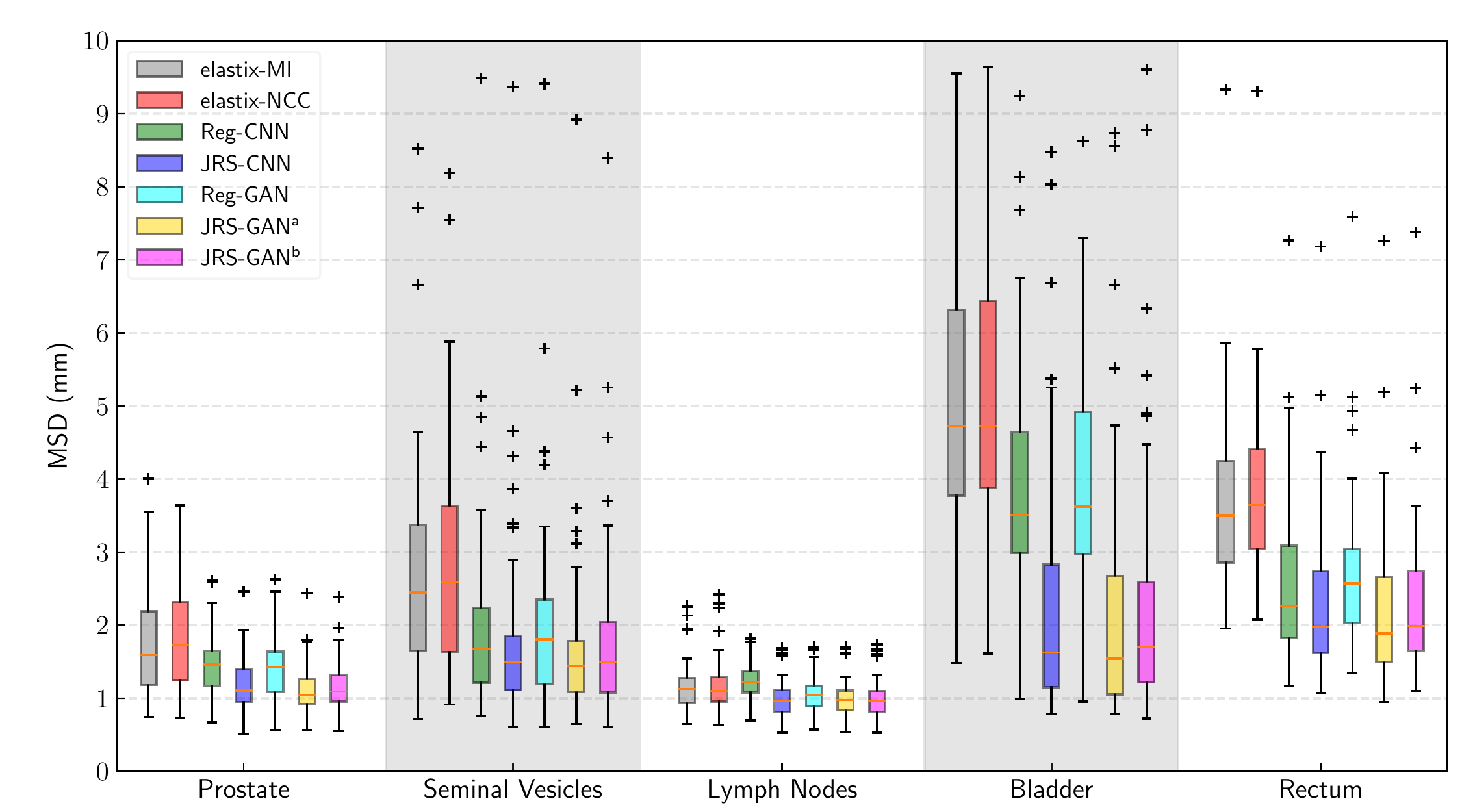}
	\caption{Boxplots for the evaluated methods in terms of MSD (mm).}
	\label{fig:boxplot}
\end{figure}

\begin{figure}[t]
	\centering
	\resizebox{\textwidth}{!}{
		\begin{tabular}{ c @{\quad} c @{\quad} c @{\quad} c}
			
			&\Huge{\texttt{elastix}-MI} & \Huge{Reg-CNN} & \Huge{JRS-GAN$^a$} \\
			\multirow{-15}{*}{\rotatebox[origin=c]{90}{\Huge{Contours}}} &
			\includegraphics[width=90mm,height=60mm]{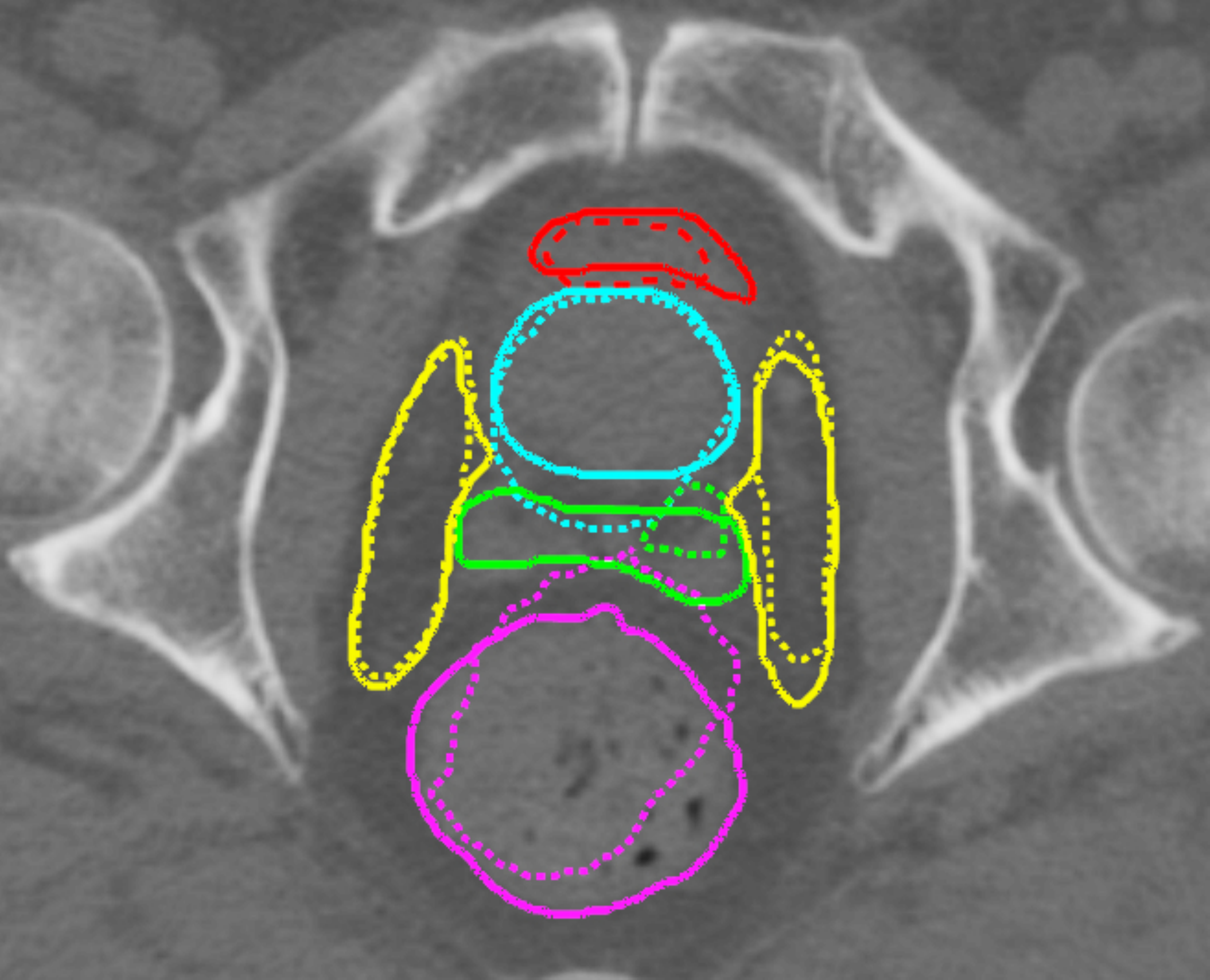} &
			\includegraphics[width=90mm,height=60mm]{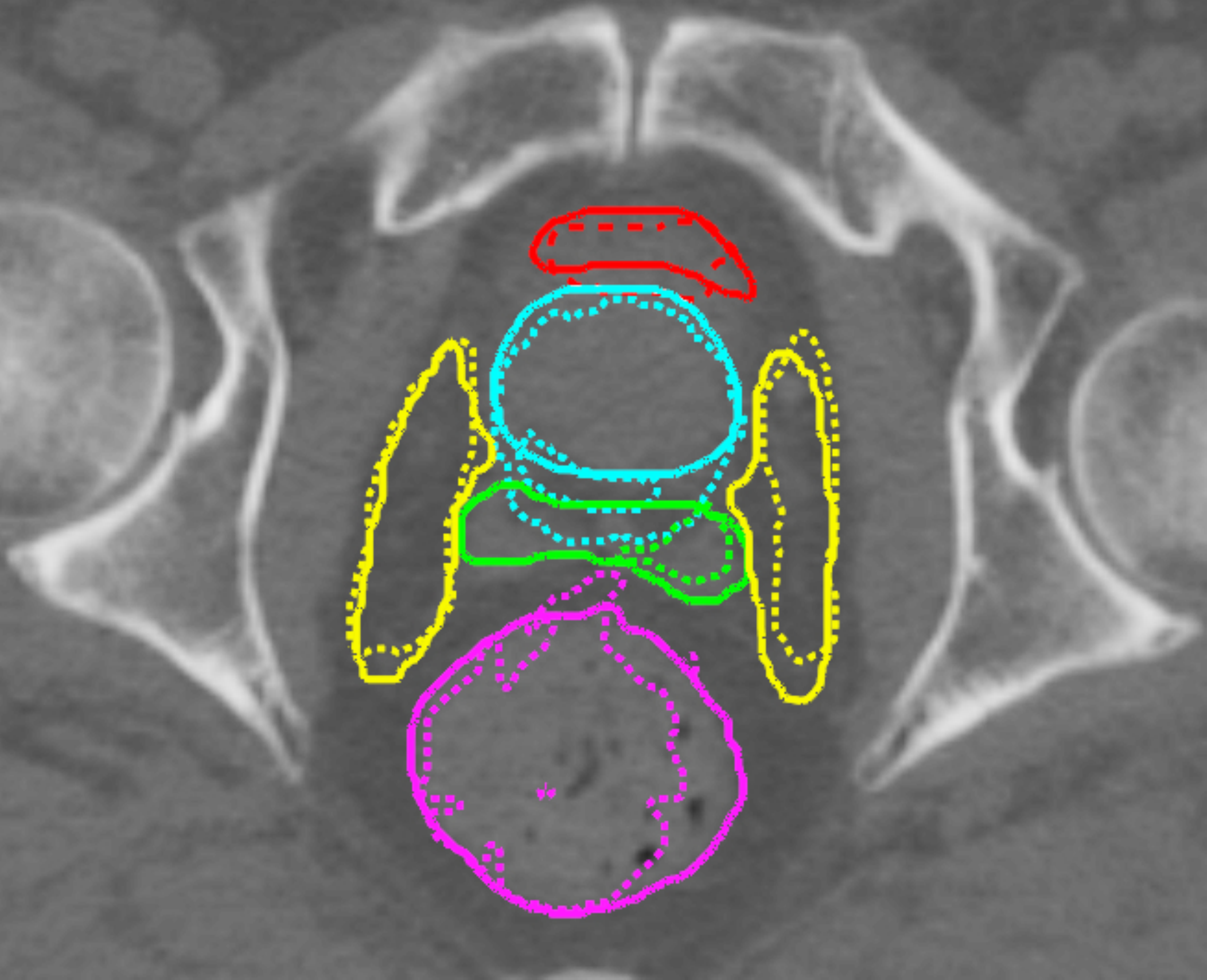} &
			\includegraphics[width=90mm,height=60mm]{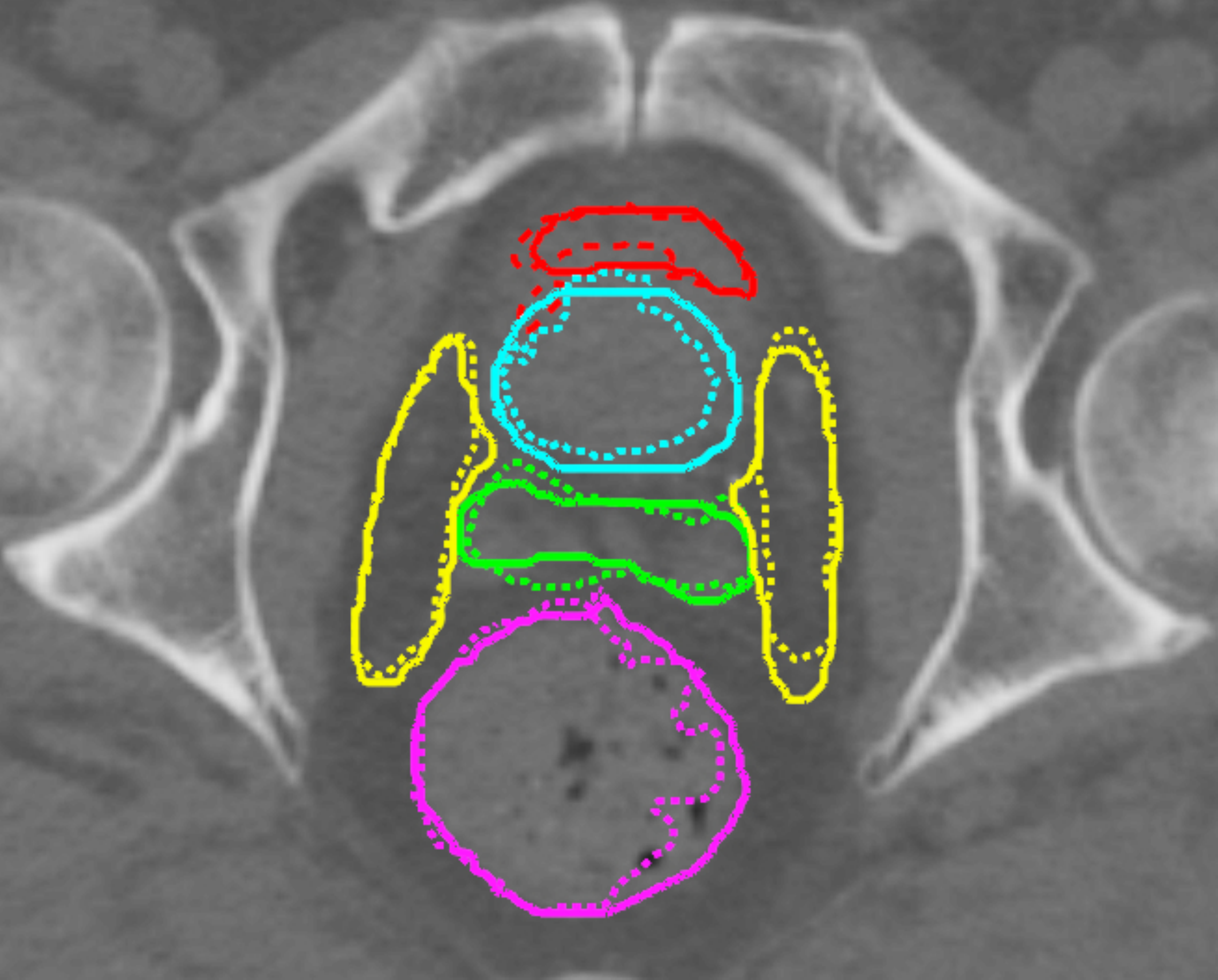} \\ \\
			
			\multirow{-16}{*}{\rotatebox[origin=c]{90}{\Huge{Abs. difference}}} &
			\includegraphics[width=90mm,height=60mm]{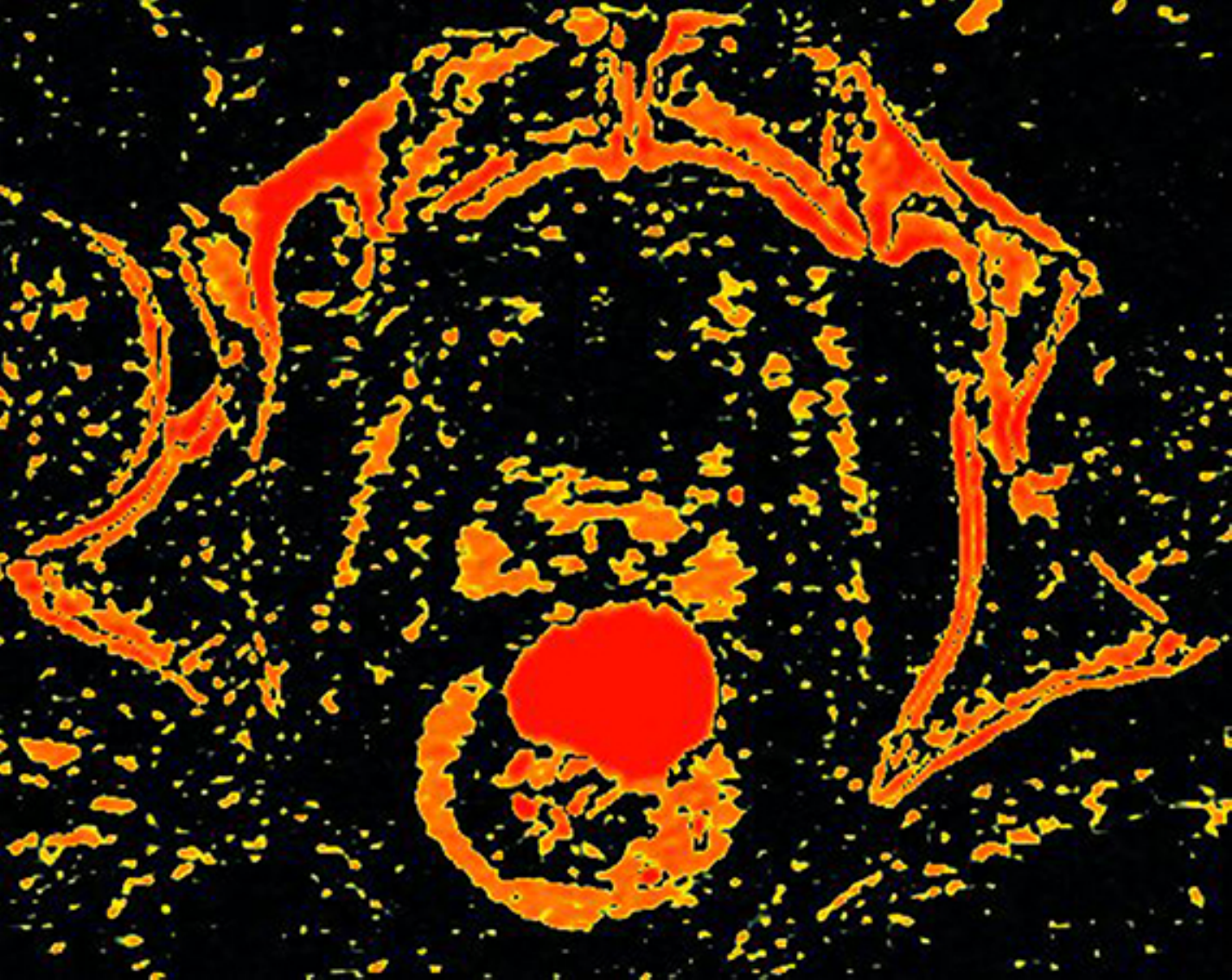} &
			\includegraphics[width=90mm,height=60mm]{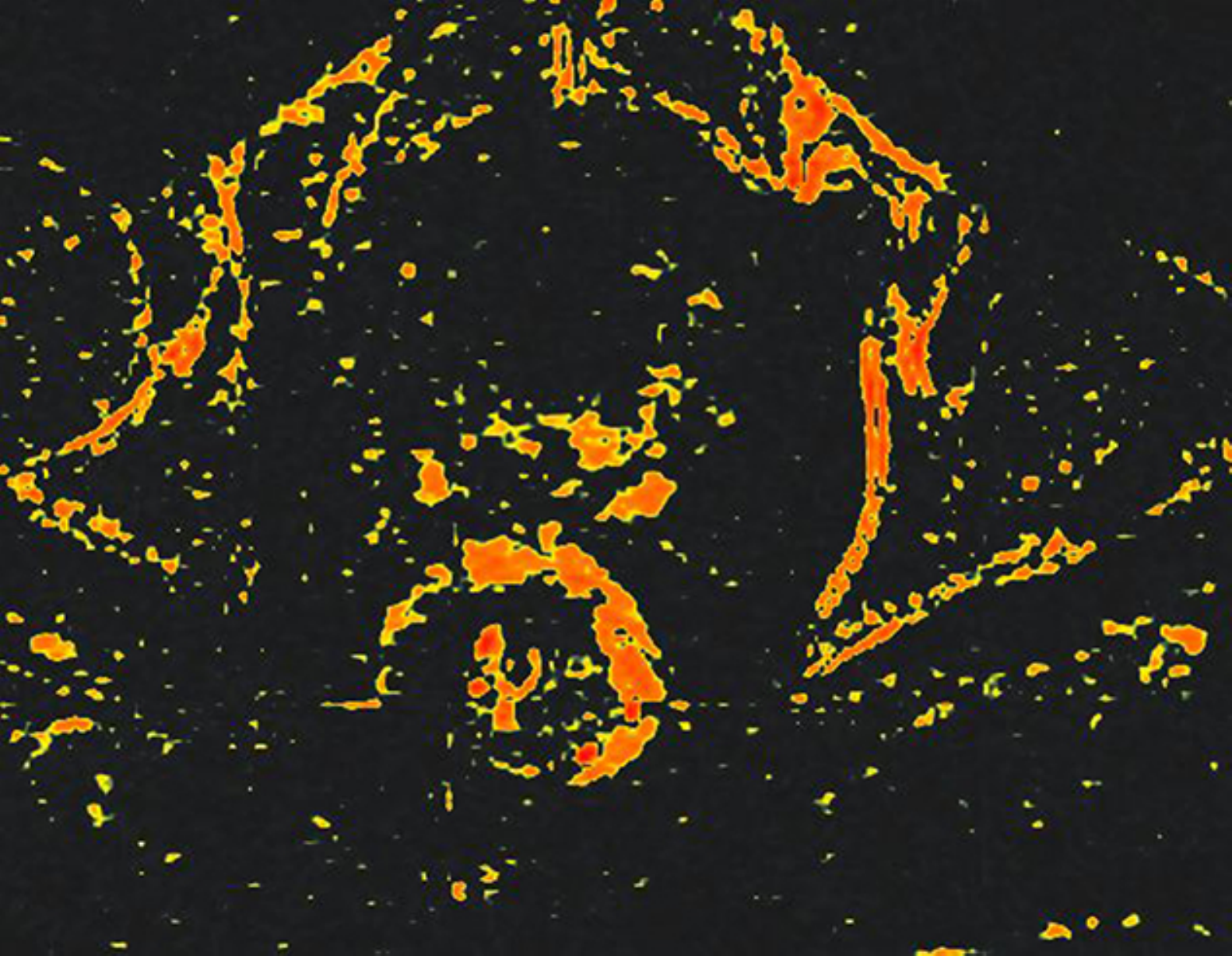} &
			\includegraphics[width=90mm,height=60mm]{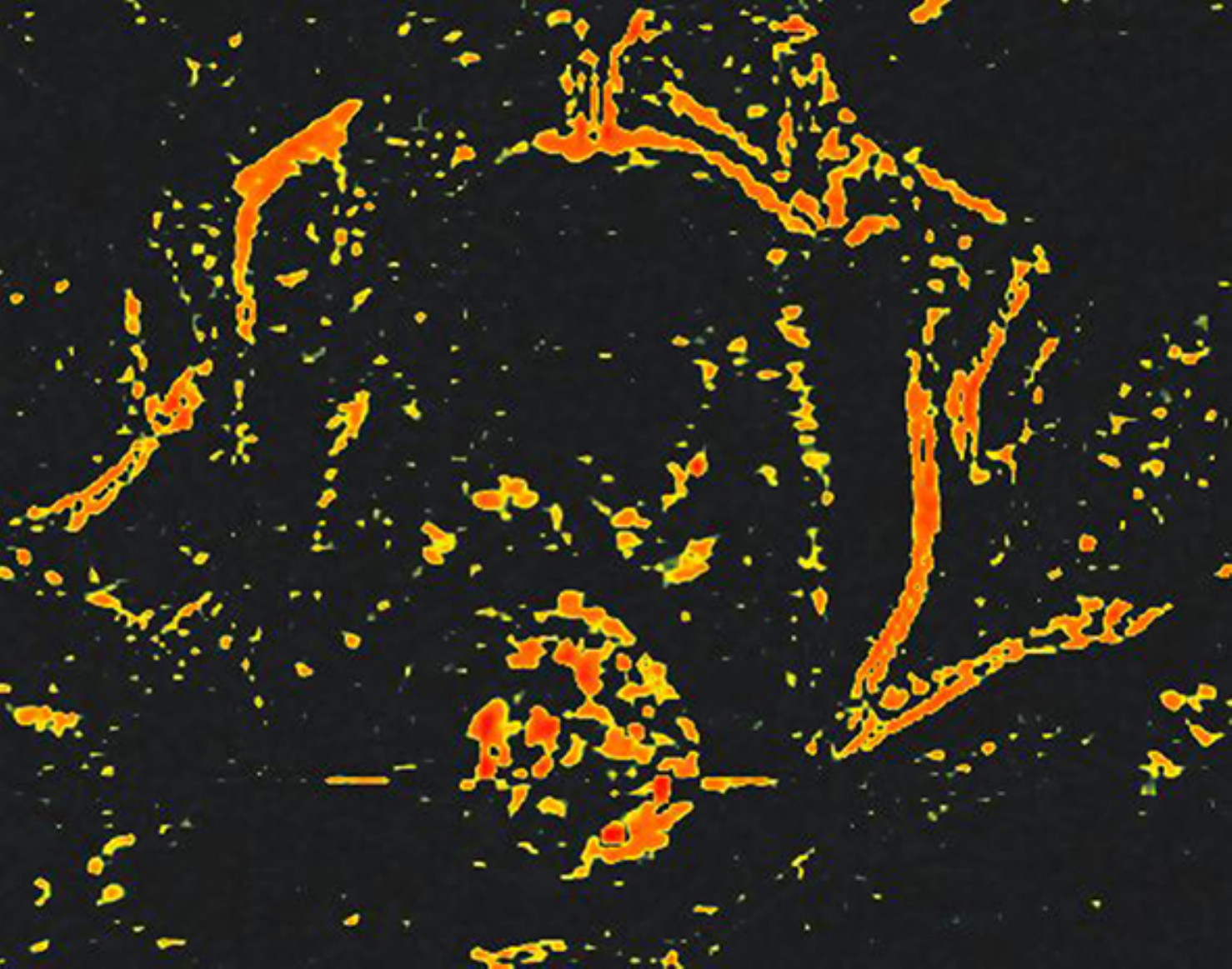} 
			
		\end{tabular}
	}
\caption{An example result for three of the methods. Top row shows the fixed image with propagated contours (solid line is manual; dotted is automatic result). The red, yellow, cyan, violet, and green contours represent the bladder, lymph nodes, prostate, rectum, and seminal vesicles, respectively. Bottom row shows heatmaps of absolute difference images between fixed and deformed moving image.}
\label{fig:examples}
\end{figure}

The MSD values in Table~\ref{table:msd} show that for all organs, the GAN-based methods significantly improved over \texttt{elastix}. This is further shown in Figure~\ref{fig:boxplot}. The results indicate a significant improvement when performing joint registration and segmentation instead of disjoint registration. Furthermore, the boxplot indicates that performance for JRS-GAN$^a$ and JRS-GAN$^b$ was very similar. Similarly, the 95\% HD values in Table~\ref{table:hd} show improvements in contour accuracy when the GAN-based method is used. Especially the organs-at-risk showed large improvements. The standard deviations of the Jacobian determinant of the estimated DVFs were $0.08 \pm 0.01$ and $0.17 \pm 0.04$ for \texttt{elastix}-MI and JRS-GAN$^a$, respectively. The average runtime for the proposed pipeline is 0.6 seconds on the GPU for a volume of size 256$^3$ voxels, while the average runtime of \texttt{elastix} at 100 iterations is 13 seconds per volume on an Intel Xeon E51620 CPU using 4 cores. Figure~\ref{fig:examples} illustrates the segmentation and registration for an example case.
 
\section{Discussion and Conclusion}
In this study, we investigated the performance of an end-to-end joint registration and segmentation network for adaptive image-guided radiotherapy. Unlike conventional registration methods, our network encodes and learns the most relevant features for joint image registration and segmentation, and exploits the combined knowledge on unseen images without segmentations.

We demonstrate that including the segmentation during training boosts the system's performance by a margin. Furthermore, adversarial feedback had a small benefit on performance, when comparing Reg-CNN with Reg-GAN. Results indicate a noticeable benefit of including segmentation masks as input to the discriminator during training. How exactly segmentation masks were embedded during training was less relevant, with only small differences observed for the seminal vesicles. This could be due to the small size and irregular nature of the seminal vesicles. A key advantage of the proposed deep learning-based contour propagation method is its runtime on new and unseen data, i.e. 0.6 s.

This work has shown that adversarial feedback can help improve registration, i.e. that a discriminator can learn a measure of image alignment. This is a promising aspect that could be further explored in future work. This will include improved GAN objectives, such as the use of gradient penalty regularization. 
 
To conclude, we have proposed a 3D adversarial network for joint image registration and segmentation with a focus on prostate CT radiotherapy. The proposed method demonstrated the effectiveness of training the registration and segmentation jointly. Moreover, it showed a substantial reduction in the computation time making it a strong candidate for online adaptive image-guided radiotherapy of prostate cancer. Since the proposed method did not only improve accuracy for the target areas, but substantially so for the organs-at-risk, this may aid reducing treatment-induced complications.

\subsubsection*{Acknowledgements.}
	This study was financially supported by Varian Medical Systems and ZonMw, the Netherlands Organization for Health Research and Development, grant number 104003012. The dataset with contours were collected at Haukeland University Hospital, Bergen, Norway and were provided to us by responsible oncologist Svein Inge Helle and physicist Liv Bolstad Hysing; they are gratefully acknowledged.

\end{document}